\documentclass[%
 preprint, 
 amsmath,amssymb,
 aps, physrev,
]{revtex4-2}

\usepackage{graphicx}
\usepackage{dcolumn}
\usepackage{bm}
\usepackage[utf8]{inputenc}
\usepackage{newunicodechar}
\newunicodechar{【}{[}\newunicodechar{】}{]}

\begin{document}

\title{\textbf{Kinetic Catalysis of Spontaneous Knotting: How Free Particles Modulate Filament Entanglement}} 

\author{Peimo Sun}
\author{Yuhan Qin}%
\affiliation{%
 Tsinglan School\\
 Songshan Lake, Dongguan, China
}%

\author{Zheng Li}
\email{Contact author: lizheng@x-institute.edu.cn}
\affiliation{
 X-Institute, Shenzhen, China
}%
\affiliation{
}%

\date{\today}

\begin{abstract}
Entangled knots form spontaneously in flexible filaments, yet the influence of the surrounding environment on this process is poorly understood. Here we demonstrate that free-moving particles act as \textbf{kinetic catalysts} for spontaneous knotting. Through controlled agitation experiments, we find that a small number of inert beads substantially enhance the probability and accelerate the rate of knot formation. This catalytic effect is non-monotonic: an optimal particle size and concentration that maximizes entanglement, while an excess of particles suppresses knotting by impeding the filament’s dynamics. We develop a stochastic model that quantitatively reproduces this behavior, attributing it to a competition between entanglement-promoting collisions and motion-suppressing drag. Our findings reveal a mechanism for tuning topological complexity, whereby adjusting these environmental agitators can either promote rapid self-assembly or inhibit unwanted entanglement. This work suggests new strategies for controlling filament topology in settings ranging from crowded biological environments to advanced materials processing.
\end{abstract}

\maketitle

Knots have been integral to human technology, from the earliest bindings of stone tools to the advanced mechanics of modern climbing ropes and surgical sutures~\cite{turner1996, patil2020topological}. This ancient concept now underpins a wide range of scientific frontiers, with topology dictating the function of DNA, the properties of polymers, the design of molecular machines and the dynamics of living matter~\cite{wasserman1986biochemical, doi1988theory, witten1989quantum, danon2017braiding, patil2023ultrafast}. However, beyond these deliberately tied or enzymatically controlled systems, spontaneous knotting of a long flexible chain represents a complex stochastic process~\cite{belmonte2001dynamic, desyatnikov2012spontaneous}. Understanding and controlling these spontaneous topological transformations is thus a key challenge with broad implications, from preventing knots in protein folding to designing novel materials~\cite{mallam2008exploring, virnau2006intricate, horner2016knot, hsu2021tying}.

The equilibrium properties of such entanglements are well described by polymer physics, where scaling laws predict that the knotting probability of a chain approaches unity with increasing length, akin to a topological phase transition~\cite{sumners1988knots, grosberg1996flory, lua2006statistics, dobay2003scaling}. However, \emph{kinetics} of this process is much less understood~\cite{tubiana2013spontaneous, orlandini2017statics}. Pioneering macroscopic experiments and simulations have quantified knotting in pristine environments, linking it to parameters such as chain length, agitation, and knot complexity~\cite{raymer2007spontaneous, ben2001knots, shimamura2002knot, vargas2017knot}. These foundational studies, often focusing on isolated chains, have been complemented by simulations exploring the effects of confinement and crowding, which suggest that environmental constraints can dramatically alter knotting statistics~\cite{arsuaga2002knotting, micheletti2006knotting, tubiana2011multiscale, shin2015kinetics, d2015molecular}. A critical gap therefore persists in the lack of systematic \emph{experimental} studies on how freely moving, non-confining particles influence knotting dynamics.

To address this gap, we investigate how freely moving particles influence the spontaneous knotting of an agitated string. Our tabletop experiments reveal that a small number of inert beads can act as \textbf{kinetic catalysts}, substantially enhancing the rate of knot formation. Strikingly, this catalytic effect is non-monotonic, with an optimal bead concentration and size that maximizes entanglement before inhibition occurs due to a crowding-induced drag. A stochastic model quantitatively explains this turnover as a balance between collisions promoting entanglement and dissipation suppressing motion. Our findings establish a new mechanism to adjust the topology of the filament and provide a tangible framework for understanding entanglement in crowded biological and industrial settings~\cite{shaw1993knotting, li2009effects}.

\textbf{Experiments}

To investigate how the stochastic knotting of a flexible line is affected by its environment, we conducted a series of tumbling experiments inspired by the seminal work of Raymer and Smith~\cite{raymer2007spontaneous}. Strings of varying lengths $L$ were placed inside a cubic container (inner side length 15cm) and agitated by rotation at a constant speed of one revolution per second for a duration of 10 s. This process was carried out under two distinct conditions: one with only the string in the container and another in which two free-moving silicone beads (each 15~mm in diameter) were introduced into the container alongside the string. After agitation, the box was opened and the final conformation of the string was classified as knotted or unknotted. All observed knot types were photographically documented (see Supplementary Materials), and further details on the experimental setup are provided in the \textit{Materials and Methods} section.

\begin{figure}[h!]
    \centering
    \includegraphics[width=0.8\textwidth]{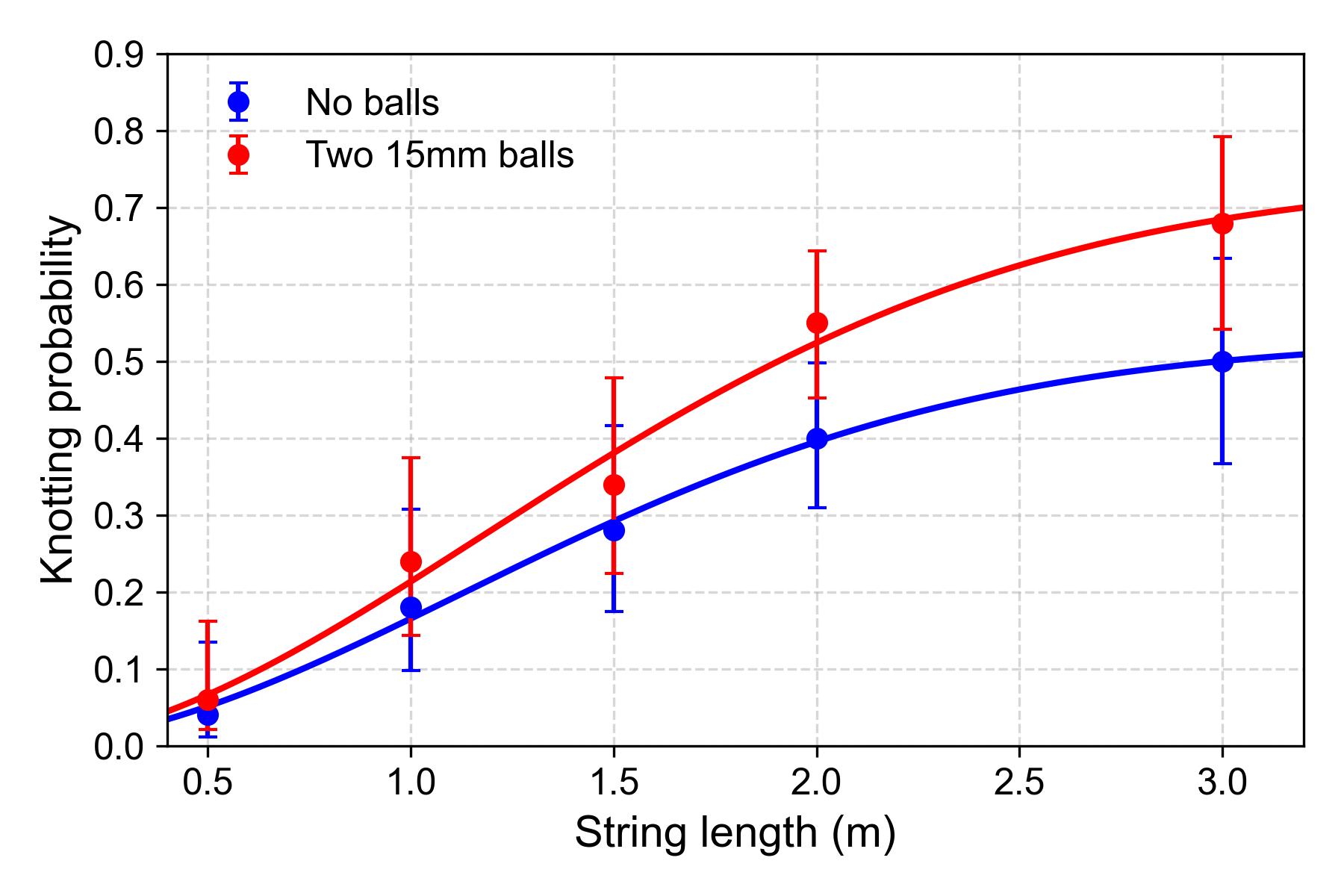} 
    \caption{Probability of spontaneous knot formation as a function of string length $L$. Data are shown for strings without any beads (blue circles) and with two 15~mm silicone beads present as free agitators in the container (red circles). Error bars represent the 95\% Wilson score confidence interval. Solid lines are fits to the kinetic model $P(L) = P_{\text{max}}\big(1 - \exp[-(L/L_c)^\alpha]\big)$, quantifying the significant enhancement in knotting probability in the presence of the beads.}
    \label{fig:1}
\end{figure}

The results indicate that the presence of these free moving beads significantly enhances the probability of knot formation in all lengths of the strings tested (Fig.\ref{fig:1}). The knotting probabilities observed for the string only condition, both in trend and in absolute value, agree well with previous findings\cite{raymer2007spontaneous}, confirming that our experimental system reproduces established baseline behavior. In particular, the addition of two beads substantially increases this probability; for example, at $L = 2.0$m, the knotting probability increases from 40\% (40/100 trials) to 55\% (55/100 trials), which corresponds to a relative increase of approximately 37\%. This result suggests that beads, even though they are not attached to the string, alter the dynamics of the system in a way that favors knotting. The solid lines in Fig.\ref{fig:1} represent fits to a kinetic model (derived and discussed below) that effectively capture the observed enhancement and enable a quantitative analysis of the underlying knotting rates.

\textbf{Discussion}

\begin{figure}[h!]
\centering
\includegraphics[width=0.9\textwidth]{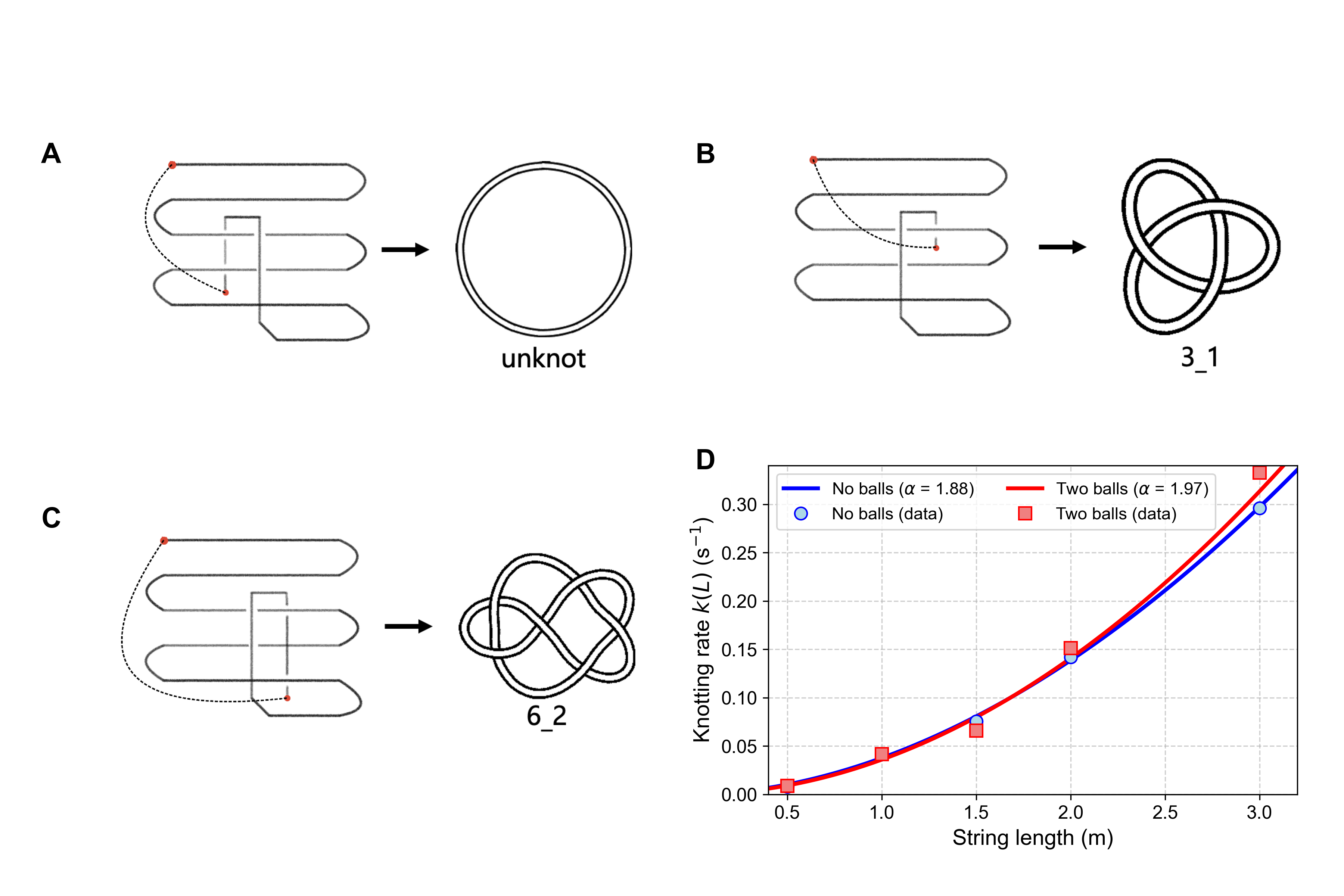}
\caption{Kinetic model of spontaneous knot formation and the resulting knotting rate. (A-C) Schematic representation of knot formation via the threading of a mobile string end (path indicated by dashed line) through loops formed by the string body. (A) A path involving multiple crossings that is topologically equivalent to the unknot. (B) A path that forms the simplest prime knot, the trefoil ($3_1$), by threading a self-generated loop. (C) A more complex path that threads a subsequent loop, resulting in a higher-order knot ($6_2$). (D) The length-dependent knotting rate, $k(L)$, derived from fitting the experimental data in Fig. \ref{fig:1} to the kinetic model.}
\label{fig:2}
\end{figure}

The kinetic principles governing the observed enhancement in knotting probability can be elucidated through a conceptual model that simplifies the complex stochastic trajectory of the agitated string. In this model, the conformation of the string is abstracted into an idealized structure consisting of multiple parallel segments (Fig.~\ref{fig:2}A–C). This representation is analogous to a mathematical braid diagram, with the crucial distinction that the segments are sections of a single continuous filament rather than separate strands. Within this framework, one end of the string is treated as a mobile terminus actively exploring the available conformational space, whereas the remainder of the string serves as a quasi-stationary network of obstacles and pathways.

The formation of a knot is contingent on a critical kinetic event: the successful 'threading' of the mobile terminus through a loop created by the string’s own body. The final topological state is highly sensitive to the precise path taken by the moving end. For instance, a trajectory involving sequential under-crossings of adjacent segments may create transient loops, but if the terminus exits without locking the topology, the conformation remains topologically trivial (reducible to the unknot via Reidemeister moves; Fig.\ref{fig:2}A). In contrast, the simplest non-trivial prime knot, the trefoil ($3_1$), is formed when the mobile end creates a loop and then threads through that same loop, creating an irreducible crossing (Fig.\ref{fig:2}B). This threading action constitutes the fundamental topological transition from unknotted to knotted state.

This model also provides an intuitive explanation for the rapid formation of highly complex knots. In three-dimensional space, the coiled conformation of the string presents the mobile terminus with a cascade of potential loops to thread. Seemingly minor deviations in the end trajectory can lead to vastly different topological outcomes. For example, if the terminus bypasses a nearby loop and instead threads a more distant one (as depicted in Fig.~\ref{fig:2}C), a knot of significantly higher complexity will be generated, such as the $6_2$ knot. This mechanism explains how even a short duration of agitation can produce intricate knots: Small-scale spatial movements of the string’s end can correspond to large-scale topological reconfigurations of the filament.

As discussed above, knot formation can be viewed as a sequence of random crossing events, and the high frequency of such events during agitation suggests that the process may be modeled as a Poisson process\cite{last2018lectures}. We therefore adopt the kinetic framework,
\begin{equation}
    P(L) = P_{\text{max}}(1 - e^{-\lambda(L)})
    \label{eq:main_model}
\end{equation}

where $P_{\text{max}}$ is the saturation probability at long length and $\lambda(L)$ represents the effective number of knotting attempts in the 10 s agitation period. The corresponding knotting rate $k(L) = \lambda(L)/T$ (with $T = 10$s) is plotted in Fig.\ref{fig:2}D. The presence of beads clearly increases $k(L)$ across the entire range of $L$. Moreover, for both experimental conditions, $k(L)$ grows faster than linearly with $L$, suggesting a scaling law behavior\cite{dobay2003scaling, tubiana2011multiscale}. This observation motivates us to model the knotting propensity with a power-law form.
\begin{equation}
    \lambda(L) = (L/L_c)^\alpha
    \label{eq:lambda_model}
\end{equation}

where $\alpha$ is a dimensionless exponent governing the knotting dynamics and $L_c$ is a characteristic length scale. Because the container geometry and the agitation protocol are identical in both conditions, we assume $L_c$ to be the same with and without beads. We determined the parameters by a simultaneous fit, with $L_c$ restricted to a single value, and the results are summarized in Table\ref{tab:fit_params}. The introduction of beads increases the scaling exponent from $\alpha \approx 1.88$ to $\alpha \approx 1.97$, indicating a direct acceleration of the knotting kinetics. In addition, $P_{\text{max}}$ is higher under the beads, consistent with the overall higher knotting probability observed in Fig.~\ref{fig:1}. 
\begin{table}[h]
\caption{\label{tab:fit_params}%
Model parameters from a constrained fit. $L_c$ was determined from the "No balls" data and held constant for the "Two balls" fit.
}
\begin{ruledtabular}
\begin{tabular}{lccc}
\textrm{Condition} & $P_{\text{max}}$ & $L_c$ (m) & $\alpha$ \\
\colrule
No balls (baseline) & 0.527 & 1.681 & 1.88 \\
Two balls (constrained) & 0.705 & 1.681 & 1.97 \\
\end{tabular}
\end{ruledtabular}
\end{table}

\begin{figure}[h!]
\centering
\includegraphics[width=0.8\textwidth]{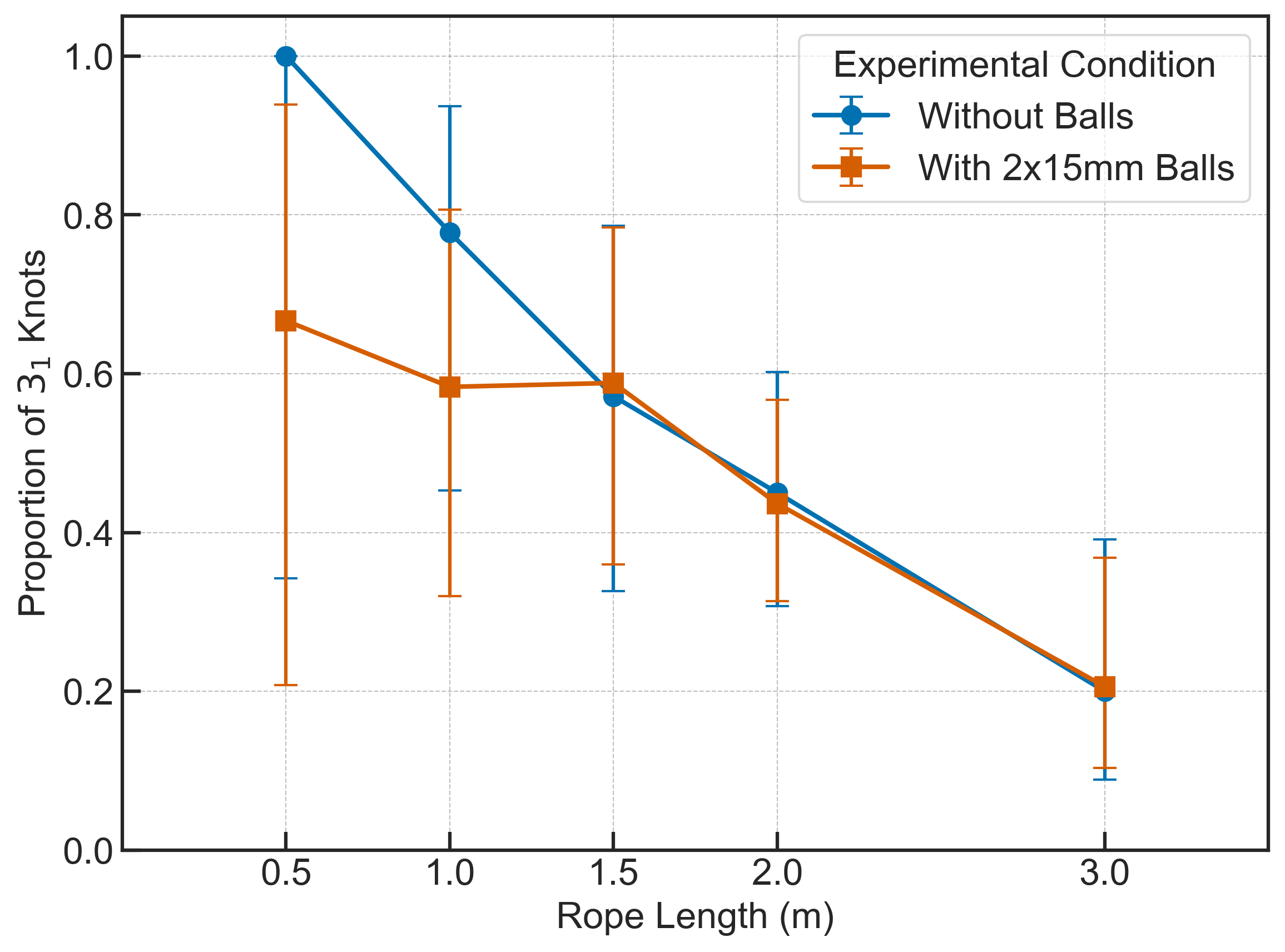} 
\caption{Proportion of trefoil ($3_1$) knots among all knotted outcomes as a function of string length. The decrease in this proportion with length indicates an increase in knot complexity. The data for conditions with and without beads are statistically indistinguishable, suggesting the beads do not alter the complexity of the knots formed.}
\label{fig:3}
\end{figure}

To further elucidate the mechanism by which beads influence knotting, we analyzed the topological complexity of knots formed under each condition. Figure~\ref{fig:3} plots the proportion of the simplest prime knot—the trefoil ($3_1$) among all knotted outcomes, as a function of string length. This proportion serves as an inverse proxy for average knot complexity: a lower fraction of trefoils indicates that, on average, more complex knots (with higher minimum crossing numbers) are being formed.

Two key features are apparent in Fig.~\ref{fig:3}. First, for both conditions, the proportion of trefoil knots decreases as $L$ increases, confirming that longer strings tend to form more complex knots. Second, and most importantly, the knot complexity distribution is statistically indistinguishable between the two conditions. In other words, the presence of beads, despite dramatically increasing the overall incidence of knotting, does not significantly change the types of knots produced.

This finding supports our kinetic model and clarifies the different roles of its parameters. The increase in knot complexity with length is a geometric effect, captured by the $(L/L_c)^\alpha$ term in Eq.(\ref{eq:lambda_model}): a longer string presents a more intricate landscape of potential loops for the end to thread (as conceptualized in Fig.\ref{fig:2}C), naturally leading to higher-order knots. In contrast, the influence of the beads is mainly kinetic, encapsulated by the change in exponent $\alpha$. The beads act as catalysts that accelerate the rate at which the string explores its conformational space, without biasing the particular topological pathway. Thus, the beads raise the knotting probability (by increasing the attempt rate) but do not preferentially favor one knot type over another. 

\textbf{Effect of Bead Number and Diameter}

\begin{figure}[h!]
\centering
\includegraphics[width=0.9\textwidth]{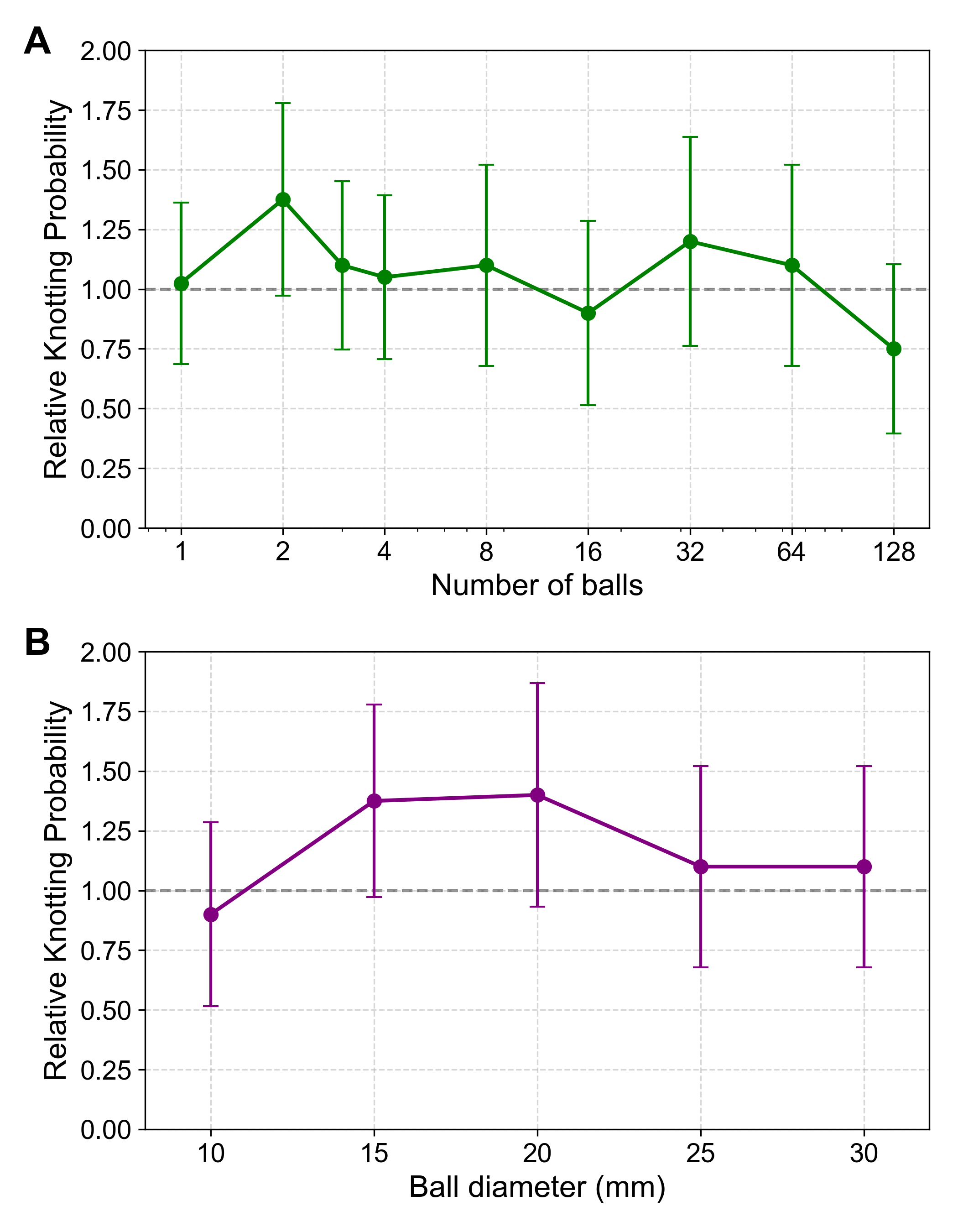} 
\caption{Non-monotonic influence of bead parameters on knotting probability. (A) Knotting probability relative to the no-bead baseline, as a function of the number of 15 mm beads in the container. (B) Relative knotting probability for two beads as a function of bead diameter. The relative probability is the ratio of the knotting probability with beads to that without beads. An optimal enhancement is observed in both cases.}
\label{fig:4}
\end{figure}

Next, we investigate how the magnitude of the bead perturbation, both in terms of bead quantity and size, affects the knotting probability. Figure~\ref{fig:4}A shows the knotting probability relative to the bead-free case as a function of the number $N$ of beads (for a fixed bead diameter of 15mm), while Fig.\ref{fig:4}B shows the relative probability for two beads of varying diameter. In both cases, the influence of the beads is \textit{ nonmonotonic}. Specifically, Fig.\ref{fig:4}A reveals an optimal enhancement around $N=2$ beads, and Fig.\ref{fig:4}B indicates that beads with diameters in the range of approximately 15–20~mm are the most effective in increasing knotting. When the beads are very large or very numerous, the knotting probability can actually be \textit{lower} than the baseline (no-bead) case.

This non-monotonic behavior can be understood through a physical picture in which an optimal number and size of beads maximize the internal motion of the end of the string relative to its body, while excessive mass or volume of beads begins to inhibit those internal movements. If the total inertial and volumetric impact of the beads is too large (as is the case for many beads or very large beads), the beads tend to co-move with the rotating container as a collective body. In effect, the string is dragged along with this motion, acting more like a rigid object and experiencing fewer internal fluctuations. This collective motion suppresses the small-scale relative movements of the string that are required at the end of the string to weave through loops and form knots. In contrast, a small number of appropriately sized beads act as discrete agitators. Their collisions with the string impart localized jostling to segments of the string, effectively increasing the frequency of crossing events between the end of the string and its body. In this optimal regime, the beads induce additional internal dynamics without imposing significant overall drag on the string.

Interestingly, this non-monotonic dependence on obstacle size and number is consistent with observations in polymer physics. Simulations by Shin \textit{et al.}\cite{shin2015kinetics} on polymer looping dynamics found an optimal crowder size that maximized loop formation rates, with smaller crowders enhancing polymer looping up to a point, beyond which further increasing crowder size or concentration led to a slowdown due to increased effective viscosity. Similarly, in our macroscopic system, beads act as inert 'crowders' that can promote or hinder entanglement depending on their total mass and size. The optimal diameter of the bead that we observe (on the order of a few centimeters, comparable to the dimension scale of the box) qualitatively matches the notion of an optimal crowder size in the microscopic system of ref.\cite{shin2015kinetics}. Thus, the beads in our experiments effectively function as kinetic catalysts that dynamically increase the rate of knotting attempts when tuned to the right size and quantity. These results suggest that the entanglement rate of filamentous materials can be controlled through carefully chosen mechanical agitation parameters and the introduction of appropriately sized particulate matter. 

\textbf{Conclusion}

In this study, we systematically examined how the presence of free-moving particles influences the spontaneous knotting of an agitated string. The results show that the introduction of a small number of silicone beads acts as a kinetic catalyst, significantly increasing the overall probability of knot formation. This enhancement is quantitatively captured by our kinetic model as an increase in the knotting-rate exponent, reflecting an accelerated rate of topological exploration, while the knot complexity remains primarily determined by the string length. We also observed that the catalytic effect of the beads is non-monotonic: there exists an optimal bead number and size that maximize the knotting probability, beyond which additional or larger beads actually suppress knot formation. This behavior is consistent with a competition between enhanced internal agitation of the string (promoting entanglements) and increased effective inertia or drag (inhibiting internal motion) when too many or too large beads are present. These results suggested that the entanglement propensity of a macroscopic filament can be actively tuned. Our findings provide a basis for understanding how the knotting rate of polymeric or rope-like materials might be modulated through mechanical agitation and the inclusion of appropriately selected inert particles.

\textbf{Materials and Methods}

The experimental apparatus consisted of a cubic box (inner side length 15cm) 3D-printed from PETG (polyethylene terephthalate glycol). The box was mounted on a computer-controlled stepper motor and rotated about a horizontal axis. The string used was a 3mm diameter braided nylon cord. Solid silicone spheres of various sizes (diameters of 10, 15, 20, 25, and 30mm) served as free-moving agitators (beads) in different trials. For each trial, the box was rotated at a constant angular velocity of 1 revolution per second for a total of 10~s. To ensure smooth initiation and termination of motion, the motor was ramped up uniformly during the first half-revolution and symmetrically ramped down during the final half-revolution. Between trials, the string was carefully untangled (if knotted) and reset to a loose, roughly straight configuration to randomize the initial condition.

\bibliography{knotting}

\end{document}